# Scanning Tunneling Microscopy Study and Nanomanipulation of Graphene-Coated Water on Mica

Kevin T. He[1,3†], Joshua D. Wood[1,2,3†], Gregory P. Doidge[1,2,3], Eric Pop[1,2,3], Joseph W. Lyding[1,3*]

[1]Department of Electrical and Computer Engineering,
[2]Micro and Nanotechnology Lab,
[3]Beckman Institute for Advanced Science and Technology,
University of Illinois at Urbana-Champaign, Urbana, IL 61801, USA.

[†]These authors contributed equally

[*]Corresponding author. E-mail: lyding@illinois.edu

**ABSTRACT**

We study interfacial water trapped between a sheet of graphene and a muscovite (mica) surface using Raman spectroscopy and ultra-high vacuum scanning tunneling microscopy (UHV-STM) at room temperature. We are able to image the graphene-water interface with atomic resolution, revealing a layered network of water trapped underneath the graphene. We identify water layer numbers with a carbon nanotube height reference. Under normal scanning conditions, the water structures remain stable. However, at greater electron energies, we are able to locally manipulate the water using the STM tip.

The interface between water and various surfaces[1,2] at room temperature has been of great interest to scientists due to its relevance in geology,[3] biology,[4] and most recently, electronics.[5,6] It has been demonstrated that water behaves very differently at an interface than it does in the bulk state, forming semi-ordered "hydration layers" close to the solid surface.[7–10] However, the exact nature of these hydration layers are still not well understood and remains the source of much controversy.[11] Recent studies utilizing AFM and other methods have made progress towards putting some of these controversies to rest,[6,11–14] but atomic-resolution imaging of the interface had not yet been achieved.

Graphene[6,15–20] has already been extensively characterized by surface imaging techniques on a variety of substrates,[21–26] but only recently has it started to see use as a template for studying other molecules,[13,27,28] Graphene is ideal for coating and trapping volatile molecules for both scanning probe microscopy[13,27,29] and electron microscopy[28] studies in that it is conductive, chemically inert, impermeable,[30] and atomically conforms to most substrates.[31] In this letter, we build upon the work performed by Xu et al.[13] and use the atomic resolution and cleanliness of the ultrahigh vacuum scanning tunneling microscope (UHV-STM) to characterize water confined between monolayer graphene and the mica surface at room temperature. Unlike previous studies of graphene on mica,[6,13,14,27,29,31,32] we use graphene grown on copper via chemical vapor deposition (CVD)[33,34] rather than graphene mechanically exfoliated from graphite.[19] While CVD graphene is inferior to exfoliated graphene in terms of carrier mobility, this drawback is offset by the ability to manufacture large, monolayer sheets and transfer them onto arbitrary substrates.[34]

Our CVD process uses a methane-to-hydrogen partial pressure ratio of 2:1, as lower ratios give higher monolayer coverage.[35,36] Previous work[33] and the supporting information give more details on our growth procedure. We transfer graphene to mica with polymethyl methylacrylate (PMMA) and use successive deionized (DI) water baths to clean the graphene films from etchant contamination. The final transfer occurs on a freshly cleaved mica surface within a DI bath in contrast to previous graphene-water-mica studies.[13,27,29] In this total water immersion, we expect there to be a high amount of water initially trapped under the graphene film. We subject the samples to 60° C heating for 5 min in air to bring the PMMA-graphene system into intimate contact with the mica, driving out most of the excess water and achieving strong graphene adhesion.[37] Wet transfers had larger area coverage than dry transfers, thereby allowing STM experiments to be conducted. Thus, the water plays a critical role in bringing the graphene and mica into contact, similar to CNT film transfer.[38] After we transfer graphene onto water-coated mica, we confirm its presence by optical imaging and spectroscopy. After loading into UHV, we degas the samples at ~650-700° C for several hours to remove surface adsorbates and contaminants.

Figure 1a gives an optical image of the STM sample with a tear in the monolayer film. Monolayer graphene on transparent mica gives ~2.3% white light absorbance per layer,[39] assisting in identifying graphene coverage. To determine whether we have trapped water under the graphene, we show high wavenumber Fourier transform infrared (FTIR) spectra on samples transferred in a final bath of $H_2O$ and $D_2O$ (99.9% purity) in Figure 1b. We subtract a reference mica signal from both the $D_2O$ and $H_2O$ transmission spectra, and then we renormalize the spectra to get absorbance information. The $H_2O$ signal is noisy, as there is no $H_2O$ IR active peak in this range. However, the $D_2O$ signal peaks around 2340 and 2360 $cm^{-1}$, corresponding to the symmetric and asymmetric stretch modes of the O–D bond.[40] There is a negligible amount of $D_2O$ adsorbed on the graphene from ambient exposure, and thus we conclude that the graphene must be trapping the $D_2O$, as seen in CNTs.[40]

It is possible that the –OD group within $D_2O$ could exchange with the interlayer –OH groups in muscovite mica. Still, we believe that this exchange is minimal in our graphene transfer, as previous work showed that this exchange within muscovite required many hours of 600°C exposure to pressurized $D_2O$ vapor.[41] These conditions are quite different than our transfer conditions. The sensitivity of IR measurements to $D_2O$ monolayers under graphene is also worth noting. Sum-frequency generation (SFG) IR spectroscopy measurements of sub-monolayer, adsorbed $D_2O$ on mica gave a O–D stretch mode at ~2375 $cm^{-1}$, demonstrating the sensitivity of IR measurements to small amounts of $D_2O$ (i.e., sub-monolayer to few-layer).[42] Thus, the spectrum given in Fig. 1b most likely originates from graphene coated, adsorbed few-layer $D_2O$ on mica. Additional experimental[43] and theoretical[44] work of $D_2O$ adsorbed on graphene show similar qualitative trends (e.g. a doubly-peaked IR spectrum around 2500 $cm^{-1}$) to our observed FTIR spectra, albeit at higher wavenumbers. We attribute this shift due to graphene induced $D_2O$ confinement.[45]

Within Figure 1c, we show point Raman spectra ($\lambda_{exc}$ = 633 nm) of graphene on mica. We transferred graphene in water and using a modified dry transfer[46] process (see the supplemental information). For the graphene-coated water on mica, we show Raman spectra before and after a UHV high temperature degas at ~650 °C. We also give Raman spectra of the bare mica for reference. All graphene spectra are monolayer, as determined by the peak height $I_{2D}/I_G$ ratio,[45] the 2D band position, and the 2D full width at half maximum (FWHM).[46] The dry transferred graphene possesses a G band at $\omega_{G,d}$ ~ 1595 $cm^{-1}$. Comparing the 2D band of the dry and wet (before degas) Raman spectra, one notes a redshift of the 2D band to $\omega_{2D,d} \sim 2647$ cm$^{-1}$ (wet transferred graphene at $\omega_{2D,b} \sim 2652$ cm$^{-1}$). Strain, either uniaxial, biaxial, or inhomogeneous, can cause a peak position shift in the G and 2D bands and increase the G band FWHM.[49,50] Thus, our Raman measurements on the wet, degassed, and dry transferred graphene films could reveal a combination of doping and strain. From the dry transferred graphene 2D band position and its FWHM (~44.8 cm$^{-1}$), we determine a tensile strain $\varepsilon \sim 0.25\%$, downshifting both the 2D and G bands. Applying this shift to the G band (averaging the contributions from the G$^-$ and G$^+$ bands) gives a $\omega_{Gd,no\text{-}strain} \sim 1597$ cm$^{-1}$, consistent with graphene on bare mica.[6] Still, graphene on bare mica[3] has a G band FWHM of ~8 cm$^{-1}$, a factor of two lower than this band's FWHM of 16.3 cm$^{-1}$. The anomalously high FWHM originates from the tensile strain as well as some inhomogeneous broadening[50] caused by wrinkles in the dry transfer process. Hence, the dry transferred graphene shows the effects of missing interfacial water on graphene on mica.

In the case of wet transfer, the PMMA/graphene stacks underwent a modified RCA clean[51] (SC-2 followed by SC-1) to eliminate adsorbed metal and organic contaminants that might dope the graphene from underneath. Both spectra are of monolayer graphene,[6,47] though the onset of the D and D' bands indicates that the degassing process induced some defects (see the supplemental information). Notably, the G band downshifts after the degas (from $\omega_{G,b} = 1597$ cm$^{-1}$ to $\omega_{G,a} = 1586$ cm$^{-1}$), showing a change in doping.[52,53] Furthermore, its full width at half maximum (FWHM) increases, implying that electron-phonon coupling is lessened by decreased doping.[53] The 2D band, however, shifts from $\omega_{2D,b} = 2651$ cm$^{-1}$ to $\omega_{2D,a} = 2666$ cm$^{-1}$ after the degas, the opposite direction of what is expected for the elimination of a p-type dopant.[53] Our analysis shows that the compressive strain required to satisfy the 2D band upshift post degas would subsequently upshift the G band, the opposite of what we observe. We give further discussion in the supplemental information.

We hold that our 2D band upshift is due to local graphene band structure modification by strongly adsorbed PMMA at defects, similar to a previous report of annealed PMMA on graphene.[54] These effects are not seen in our STM measurements but are observed in the Raman measurements, as each method has different fundamental length scales. As discussed in the supplemental information, the quasi-parabolic band structure of the PMMA/graphene decreases the Fermi velocity, thereby blue-shifting the 2D band strongly and barely modifying the G band.[55] Furthermore, the invariance of the peak height $I_{2D}/I_G$ ratio before and after the degas suggests that we have not introduced additional dopants in our processing.[53] Thus, the post-degas Raman point spectrum is characteristic of CVD graphene on water on mica. Still, we provide spatial mapping to strengthen this conclusion further.

Figure 1d gives a histogram of the G band position before the degas, a Gaussian distribution centered at 1596 cm$^{-1}$ (population mean of $\omega_{G,b}$ = 1595.0±8.9 cm$^{-1}$, $n$ = 89). A previous report[6] showed that the G band for graphene on bare mica is around $\omega_G$ ~ 1595 cm$^{-1}$. Despite the similarity in G band position, we hold that many layers of water are encapsulated by the graphene during water-based transfer, as shown in Figure 1b. The introduction of this water, combined with its stability on mica,[56] makes it unlikely that we have graphene on bare mica during our Raman measurement. Before the degas in UHV, we find that STM imaging of the surface is unstable, which we attribute to adsorbed contaminants. Therefore, the high value of the G band position likely originates from remaining p-type PMMA residue[57] from the graphene transfer. It is also possible that the many layers of water possess more residual dopants, shifting the G band. Doping effects are also present in other Raman metrics (see the supplemental information).

After the ~650 °C degas, the G band's position shifts to $\omega_G$ ~ 1586 cm$^{-1}$ (population mean of $\omega_{G,a}$ = 1585.9±4.4 cm$^{-1}$, $n$ = 129), as shown in the histogram of Figure 1e. The band's position is close to previous Raman measurements[6] for graphene on single-layer water on mica ($\omega_G$ ~ 1583 cm$^{-1}$). Based on earlier reports for annealed CVD graphene (in UHV[57] and in air[54]), it appears that the high temperature degas removed most of the adsorbed PMMA residue from the graphene, downshifting the G band. The $\Delta\omega_G$ ~ 3 cm$^{-1}$ upshift between our mean G band position and the previously published work could be a sampling effect or could be attributed to p-type atmospheric adsorbates[53] and some remaining PMMA[51] within the Raman spot. Only a few points within the Raman map composing Figure 1e (see supporting information for the map) are near what is expected for graphene on bare mica, $\omega_{G,m}$ ~ 1595 cm$^{-1}$, supporting the conclusion that the graphene is covering a full, multi-layered water film. The G band's lower position is due to the water screening interfacial charge transfer[6] between the graphene and heavily p-type mica. If graphene were p-type doped by the bare mica, we would expect a strong shift in the graphene Fermi level in scanning tunneling spectroscopy (STS) measurements. We do not see this, which we discuss in the supporting information.

Scrutinizing the G band FWHM carefully raises the concern of inhomogeneous broadening[50] in the Raman spot. The large spatial sampling over which the data in Figures 1c and 1d is collected makes it unlikely that the downshift in the G band and its broadened FWHM result from inhomogeneous broadening. However, if the thermal degas introduces wrinkles into the graphene, on a scale larger than the STM images but smaller than the Raman spot, inhomogeneous broadening could occur, thereby increasing the G band FWHM. Thermally induced wrinkles in graphene and their effects on Raman were previously studied,[58] making this outcome feasible. However, we believe that doping is the dominant effect for the trends observed, but we cannot rule out inhomogeneous broadening entirely.

In Figure 2, we show a 30 nm by 30 nm STM topographic image of a typical sample surface (Figure 2a), and a zoomed-in spatial derivative (Figure 2b) illustrating the honeycomb lattice of the monolayer graphene covering. We present a larger 100 nm by 100 nm false-colored STM topograph in Figure 3c, which gives a better overview of our surface and shows the relative heights of the different features. There are three distinct water layers visible, as well as a graphene grain boundary and some taller protrusions extending from the top water layer. The presence of the grain boundary is not surprising, as CVD graphene is known to be polycrystalline,[59,60] but it is interesting to note that the water does not appear to preferentially congregate along the boundary. In light of recent AFM data suggesting that adsorbed water prefers to form droplets instead of layers centered on defects on hydrophobic surfaces,[29] we can conclude that the hydrophobicity of the CVD graphene covering has little effect on the underlying water structure.

It is possible that our high temperature degas in UHV induces strain in the graphene as the water escapes, which could deform the graphene[61,62] and influence the water structure that we observe. However, a recent AFM study demonstrated that water easily escapes from the edges of the graphene-mica interface,[14] which would imply that most of the volatile water would have already escaped during the pump-down (0% relative humidity) process before degas. Also, the presence of intact low-angle grain boundaries[63] suggests that the remaining water does not exert enough pressure when heated to seriously damage the graphene. We do not notice any major changes in the surface structure for degas times ranging from 5 hours to 30 hours. Temperature-induced stress deformities are generally large-scale wrinkles and should not affect the small surface features that we observe, such as the protrusions out of the top water layer.[62] The protrusions range from several angstroms to over one nanometer tall, and they only appear on the second or third water layer. This implies that their formation is dependent on the underlying water structure rather than on the graphene coating. A more likely explanation for these protrusions would be that they are water-surrounded contaminants or perhaps nano-droplets that have nucleated out of defects in the mica. They could also be additional layers of water which have started to exhibit bulk-like behavior due to their increasing distance from the mica surface. Molecular dynamics simulations and x-ray reflectivity data have indicated that water layers on mica cease to be easily distinguishable starting at around 1 nm away from the mica surface.[56,64] The water structures are also extremely stable over the course of our experimental observation (several days for some areas), regardless of the water layer or protrusion size.

We measure the exact number of trapped water layers by first sandwiching single-walled carbon nanotubes (SWCNTs) between the graphene and mica. The SWCNTs are deposited onto the mica via *ex-situ* dry contact transfer[65] (DCT) before the graphene covering is applied. The mica is heated during DCT to ensure that any water is removed and the SWCNTs come into direct contact with the mica surface. We use HiPco SWCNTs with very narrow diameter distribution centered on 1 nm,[66] which means that we can use the measured height of these nanotubes to extract the number of water layers. A STM topograph of a water-immersed SWCNT sandwiched between graphene and mica is shown in Figure 3a. Only part of the SWCNT is shown in the 43 nm by 43 nm scan; the total length of the nanotube is approximately 100 nm. There is a monolayer of water trapped between the SWCNT and the graphene coating, and this layer is removed using the STM tip before the height measurements are taken. More detail on this process can be found in the supplemental material. Figure 3b shows a height profile taken at the dashed red line marked in Figure 3a. The height of the second water layer is measured to be ~3 Å and the difference in height between the SWCNT and the first water layer is ~6 Å. Due to convolution with the tip geometry, the measured width of the SWCNT appears much broader than it actually is, but the height is unaffected by tip convolution and is a good gauge of the actual nanotube dimensions. Figure 3c shows a cartoon illustrating the different layer dimensions. The dotted blue arrows represent measured dimensions (2nd water layer height, difference in CNT height), the solid black arrows represent known dimensions (graphene height, total CNT height), and the dashed red arrows represent the calculated dimensions (1st water layer height). Taking the difference between the measured height of the SWCNT (~6 Å) and the known height of the SWCNT (~10 Å), we can calculate the height of the water layer, which turns out to be ~4 Å. This means that there is only 4 Å of water between the bottom layer of the image that we show in Figure 4a and the mica surface. This corresponds to approximately one layer of water and matches well with previous AFM data.[6,13]

In Figure 4, we present some statistics on the height and roughness of the water layers that we have sampled. These histograms include data from different regions on the same sample as well as data from several different samples. Figure 4a shows the height distribution of the second water layer. The heights are spread over a wide range (average of 3.5 Å), suggesting that this layer does not have a definite crystal structure. This observation is further corroborated in Figure 4b, which shows the roughness distribution of the water layers. The roughness of the second water layer again has a very wide range, suggestive of an amorphous structure. In contrast, the roughness of the first water layer is narrowly distributed and centered around 15 pm, similar to previous AFM measurements.[13]

To further explore the nature of the water trapped under the graphene monolayer, we attempt to manipulate the surface by standard STM nanolithography techniques.[67–69] Prior work demonstrated that water films on mica could be perturbed using an AFM tip[70] at room temperature, though such manipulation has not been demonstrated with a graphene coating. STM manipulation of water films at room temperature had not been possible until now, but manipulation of water at cryogenic temperatures had been previously reported.[71–74]

Figure 5 shows the creation of local pinholes in the amorphous second and third water layers. Like the non-modified water, the induced pinholes are also extremely stable over time. The topographs and associated height contours show that the created pinholes penetrate all the way through to the water layer below while leaving the graphene undamaged. The size of these pinholes can be partially controlled by adjusting the electron dose and bias potential, though their shapes tend to be non-uniform and somewhat random. We are able to manipulate the water layer at both positive (Figures 5a and c) and negative (Figures 5b and d) sample bias, whereas existing work only report successful manipulation at positive sample bias.[71–74] Of course, all previous STM manipulation work has been performed on metal substrates, where it is hypothesized that the metal surface states mediate the excitation of the water,[72,75] so it is likely that our mechanism for manipulation is quite different.

The locality of the patterns, even through bilayer graphene, implies that the tunneling electrons are bypassing the graphene coating and directly interacting with the water hydrogen bonds. The non-uniformity and randomness of the patterns also suggest that the electrons are traveling a small distance through the water after injection. The fact that we observe water manipulation at both positive and negative sample bias rules out an electric field effect, since the water always moves away from the tip, independent of field direction. Attempts to move the tip closer to the surface under zero bias showed us that manipulation did not occur as a result of the tip pushing into the water layer. Inelastic electron tunneling (IET) into the amorphous water layer does not explain the non-uniformity and the tendril-like spreading of the patterns, as all of the patterning should be localized to right under the tip apex. It is possible that the tendril-like patterns are being created by joule heating as the tunneling electrons dissipate through the water layer. The exact effect that the graphene has on these tunneling electrons as well as the states that these electrons are using is not obvious from our current data, and it will be the subject of a future systematic study.

Similar to previous AFM work,[13] we are unable to manipulate the first layer of water. This is most likely due to its crystalline structure and its strong adherence to the hydrophilic mica surface. However, we do not believe the crystalline structure of the first water layer to be ice Ih, as previously claimed.[13] Ice Ih has a hexagonal lattice structure, which should form a hexagonal moiré pattern with the graphene lattice, depending on their relative alignment. We have imaged many different graphene orientations over the course of our experiments, but have never observed a moiré pattern exclusive to the first water layer. The hexagonal moiré patterns that we did observe were due to the presence of stacked graphene and were visible over all the water layers (see supporting information).

A possible explanation for the structure of the first water layer is that while it does not have a well-defined, periodic crystal structure, it is strongly bound to the mica surface. The hydration layer on mica has been the subject of many theoretical[64,76] and experimental studies,[8,10,56] though its exact thickness and behavior are still contested.[9,11] From our data, as well as previous research,[6,12,13,56,64] we argue that the thickness of the hydration layer on mica is ~1 nm, and is split into three distinct water layers. The first water layer is strongly bound to the mica surface, with a thickness of ~4 Å. This layer cannot be manipulated, and exhibits properties similar to a crystalline solid. The second and third water layers, on the other hand, while still more viscous than bulk water, are much more amenable to manipulation than the first layer. They are stable in equilibrium at room temperature, but high tunneling conditions can break bonds and cause them to rearrange. Beyond layer three, the water begins to exhibit bulk-like behavior as the layers start to blend together.

In summary, we performed UHV-STM at room temperature on few-layered water trapped between monolayer graphene and mica. The graphene coating keeps the water stable on the surface and protects it from high temperature processing in vacuum, but does not otherwise perturb or alter the water bonding structure, even at the higher defect-density grain boundaries. We observe up to three layers of water trapped between the graphene and mica, with the first layer being strongly bound while the second and third layers are amorphous. We also demonstrate the ability to manipulate the amorphous water layers using the STM tip. This work demonstrates the feasibility of using CVD graphene coatings for nano-templating in high resolution STM studies as well as furthering our understanding of water behavior near the mica surface. Graphene-coated water will allow further STM-based research of other aqueous suspended structures, such as biomolecules in water.

**Acknowledgment**

The authors graciously acknowledge Dr. Gregory Scott, Marcus Tuttle, and Prof. Martin Gruebele for assistance with FTIR measurements and enlightening discussion. We thank Feng Xiong for help in AFM experiments. We also acknowledge Justin Koepke for useful data on CNT diameter distributions. This work was supported by the Office of Naval Research under grants N00014-06-10120 and N00014-09-1-0180, and the National Defense Science and Engineering Graduate Fellowship through the Army Research Office (JDW).

**Supporting Material**

We include the following: experimental methods; optical microscopy images; additional Raman spectra and analysis; STM images of turbostratic bilayer graphene; scanning tunneling spectroscopy (STS) data for graphene on water on mica; the procedure used for height and roughness analysis of our water layers; and graphene grain boundary imaging by STM. This material is available free of charge via the Internet at http://pubs.acs.org.

**Figures**

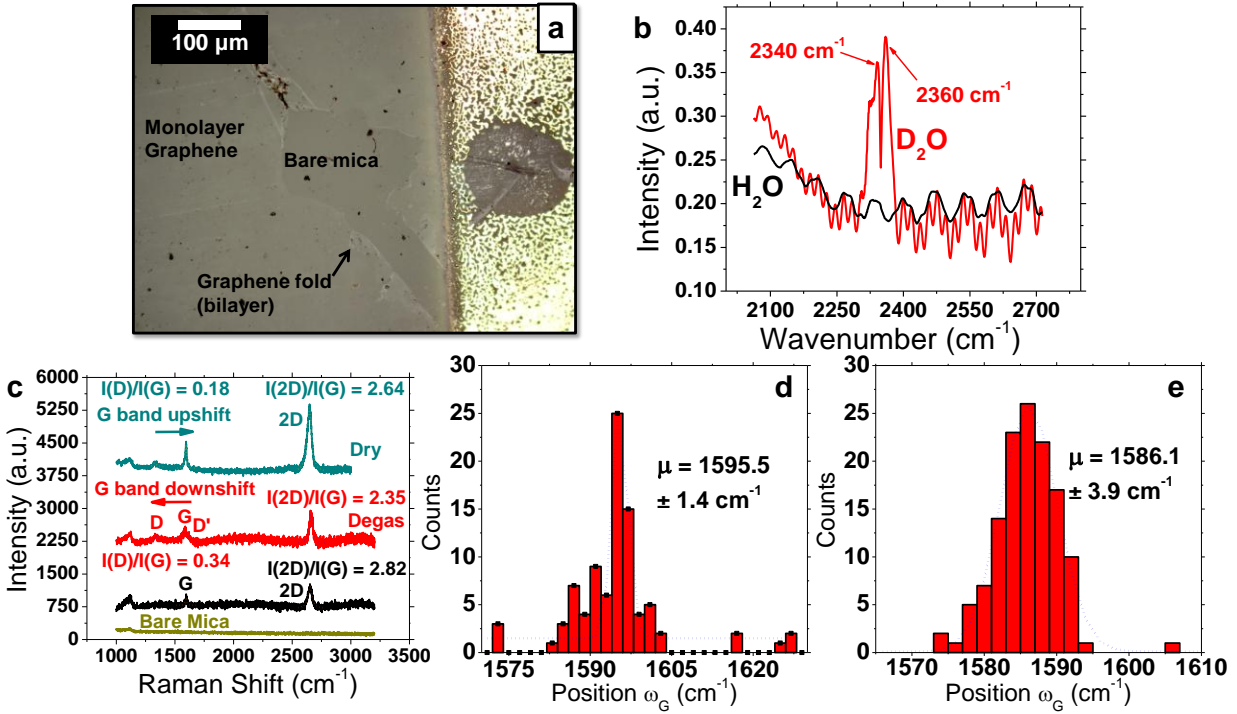

**Figure 1.** Optical characterization and spectroscopy of graphene-coated water on mica. **(a)** Optical image of the contacted sample used in STM experiments, showing monolayer graphene, folds in the CVD film, and the bare mica through a tear in the graphene. **(b)** Fourier transform infrared (FTIR) spectra of graphene transferred to mica in final baths composed of $H_2O$ and $D_2O$ showing a doubly-peaked signal for trapped $D_2O$ under graphene. This is contrasts with the trapped $H_2O$ signal, which is simply noise. Both peaks correspond to stretch modes for the O-D bond, confirming the heavy water trapped by graphene. **(c)** Point Raman spectra ($\lambda_{exc}$ = 633 nm) of dry transferred monolayer graphene (intensity ratio $I_{2D}/I_G > 2$ from peak fitting) on mica and $H_2O$-transferred graphene before (black) and after (red) a high temperature degas. The dry transferred graphene's G band position is upshifted to ~1595 cm$^{-1}$, whereas the degas introduces some defects and downshifts the G band to ~1586 cm$^{-1}$ for trapped few-layer water. Histogram of G band position from Raman mapping before **(d)** and after **(e)** the ~ 650 °C degas. After the degas, the G band's mean position is close to what is expected for graphene-coated, few-layer water on mica.

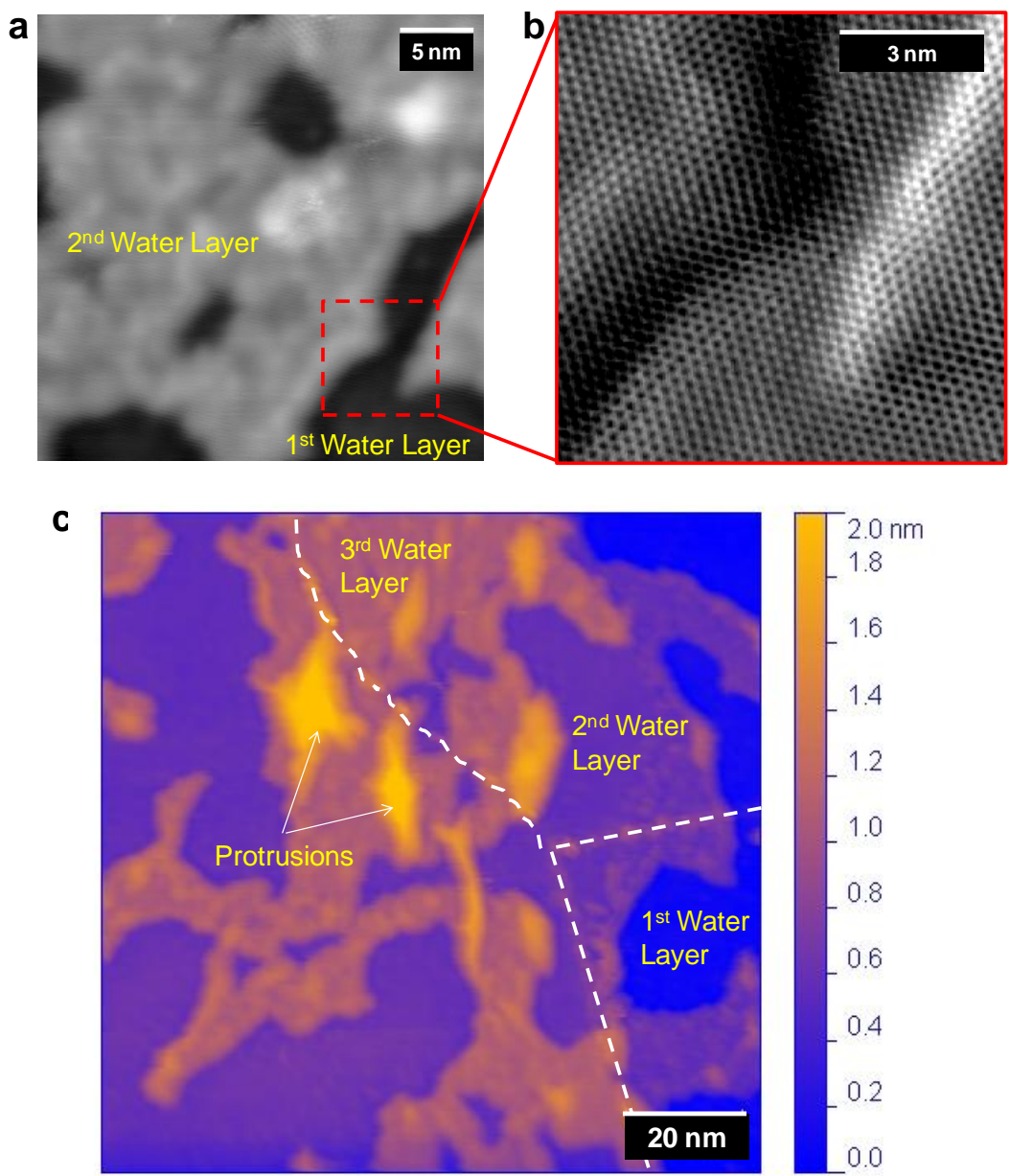

**Figure 2.** Scanning tunneling microscopy topographic scans of few-layered water confined between graphene and mica. **(a)** 30 nm by 30 nm image showing the first two water layers on the mica surface. **(b)** Zoomed-in spatial derivative of the boxed region in (a) showing the honeycomb lattice of the monolayer graphene coating. **(c)** 100 nm by 100 nm false-colored topographic image of graphene-water-mica system. Three layers of water are visible, as well as a graphene grain boundary, which is labeled by the dotted white line. The protrusions coming out of the third water layer could be due to either contaminants trapped under the graphene, or to the water displaying increasing bulk-like properties as it gets further from the mica surface. Scanning conditions are –0.35 V sample bias and 1 nA tunneling current.

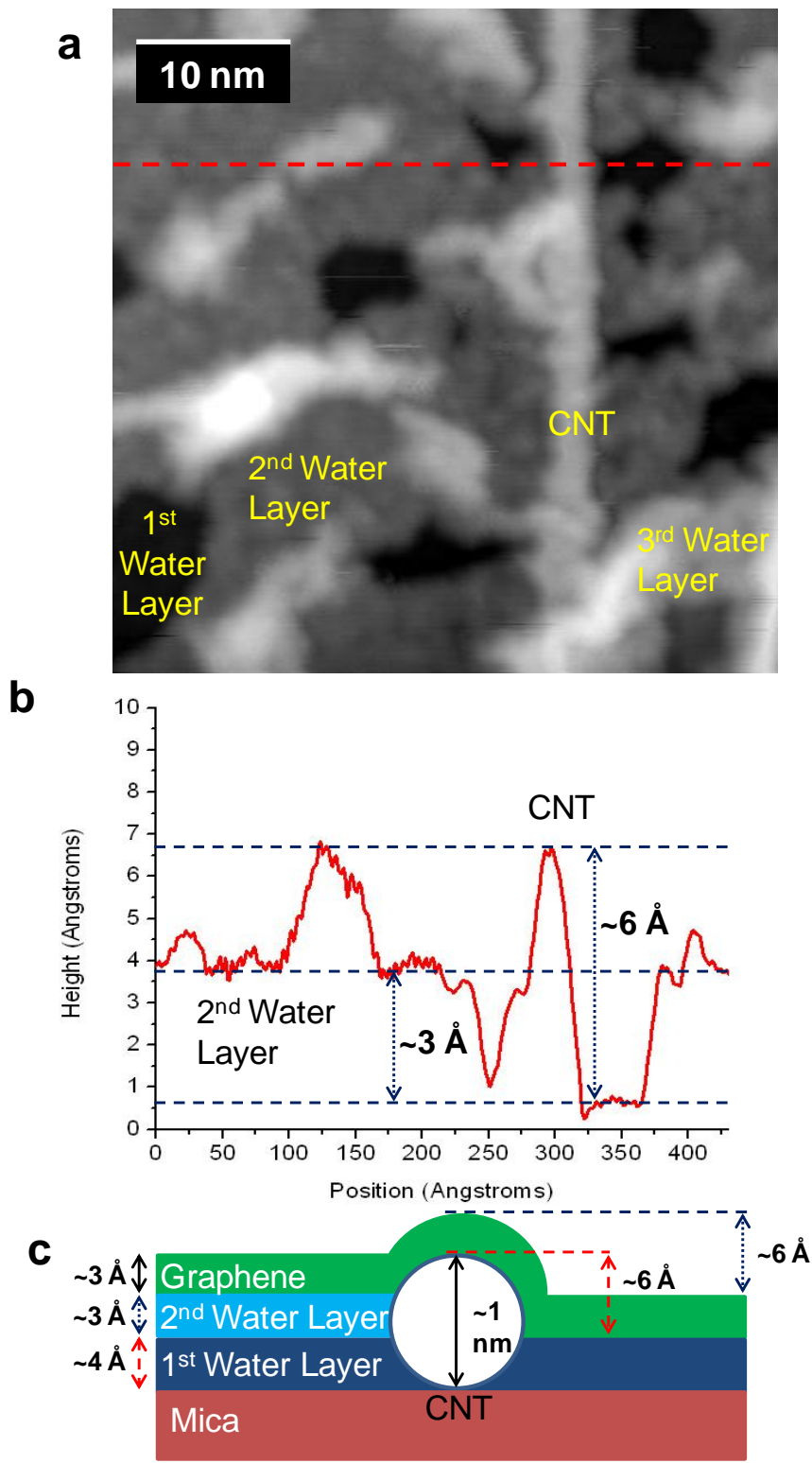

**Figure 3.** (a) 43 nm by 43 nm topographic STM image of a single-walled carbon nanotube embedded in the confined water layers between the graphene and mica. The first and second water layers are clearly defined, while the sporadic clusters appear to be the beginnings of a third

water layer. **(b)** Height profile taken at the dotted red line in (a). Here, the second water layer appears to be approximately 3 Å tall, while the SWCNT juts 6 Å above the first water layer. **(c)** Cartoon showing how we determine the heights of each of the water layers in this image. The dotted blue arrows are the values that we measured in (b): 3 Å for the second water layer and 6 Å for the part of the SWCNT above the first water layer. The black arrows are the heights that we know from external references: ~3 Å in height for monolayer graphene and ~1 nm for our HiPco SWCNTs. The red arrows represent the heights that we derived from our known quantities. Knowing the total height (~1 nm) of our SWCNT and how much it juts out of the first water layer (~0.6 Å), we can subtract and determine that there is indeed only one layer of water between the graphene and mica, and that the height of this layer is ~4 Å. Scanning conditions were -0.35 V sample bias and 1 nA tunneling current.

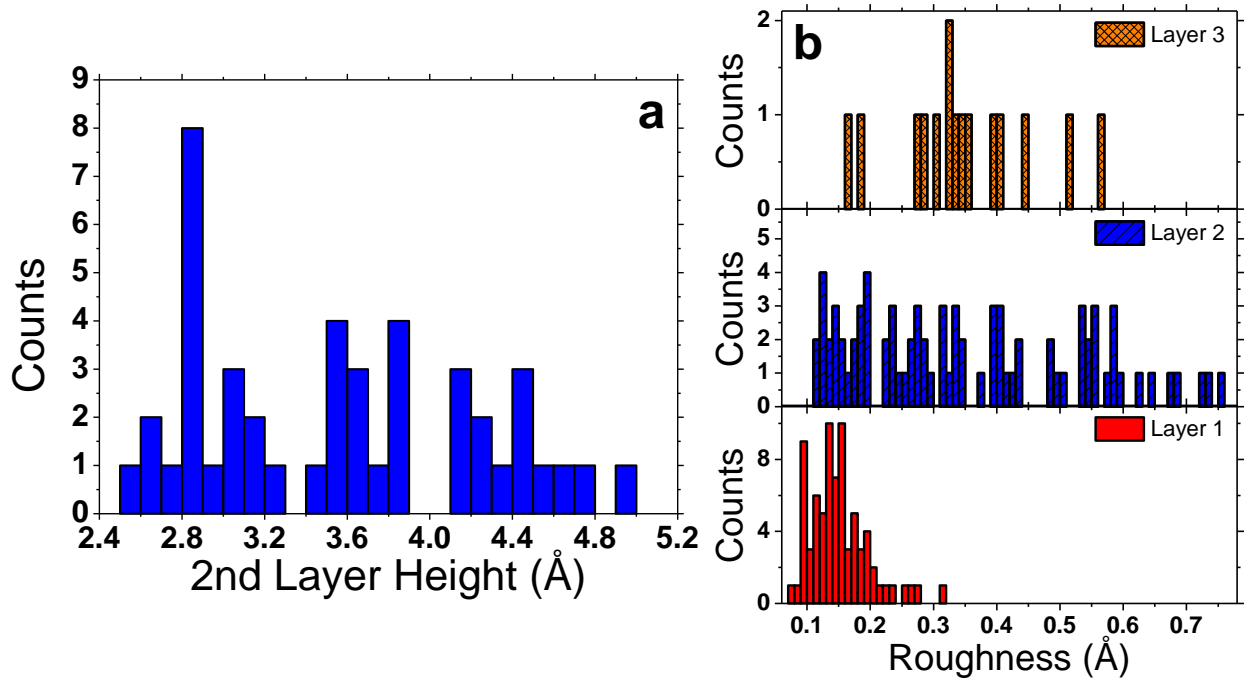

**Figure 4. (a)** Histogram of the height distribution of the second water layer. The data for this histogram was collected from four different samples, though each sample was prepared in a similar fashion. The average height is 3.5 Å, though the spread is quite large, and there is no clear trend. **(b)** Histogram of the roughness distribution for the three water layers that we have observed. This data was collected from the same four samples as the height measurements. We see that the roughness distribution of the first water layer is fairly narrow and centered at approximately 15 pm, similar to AFM measurements reported previously. The roughness distribution for the second and third water layers, however, similar to the height distribution of the second water layer, is very spread out without a clear trend. This suggests that while the first layer may have a more well-defined structure, the second and third layers are amorphous.

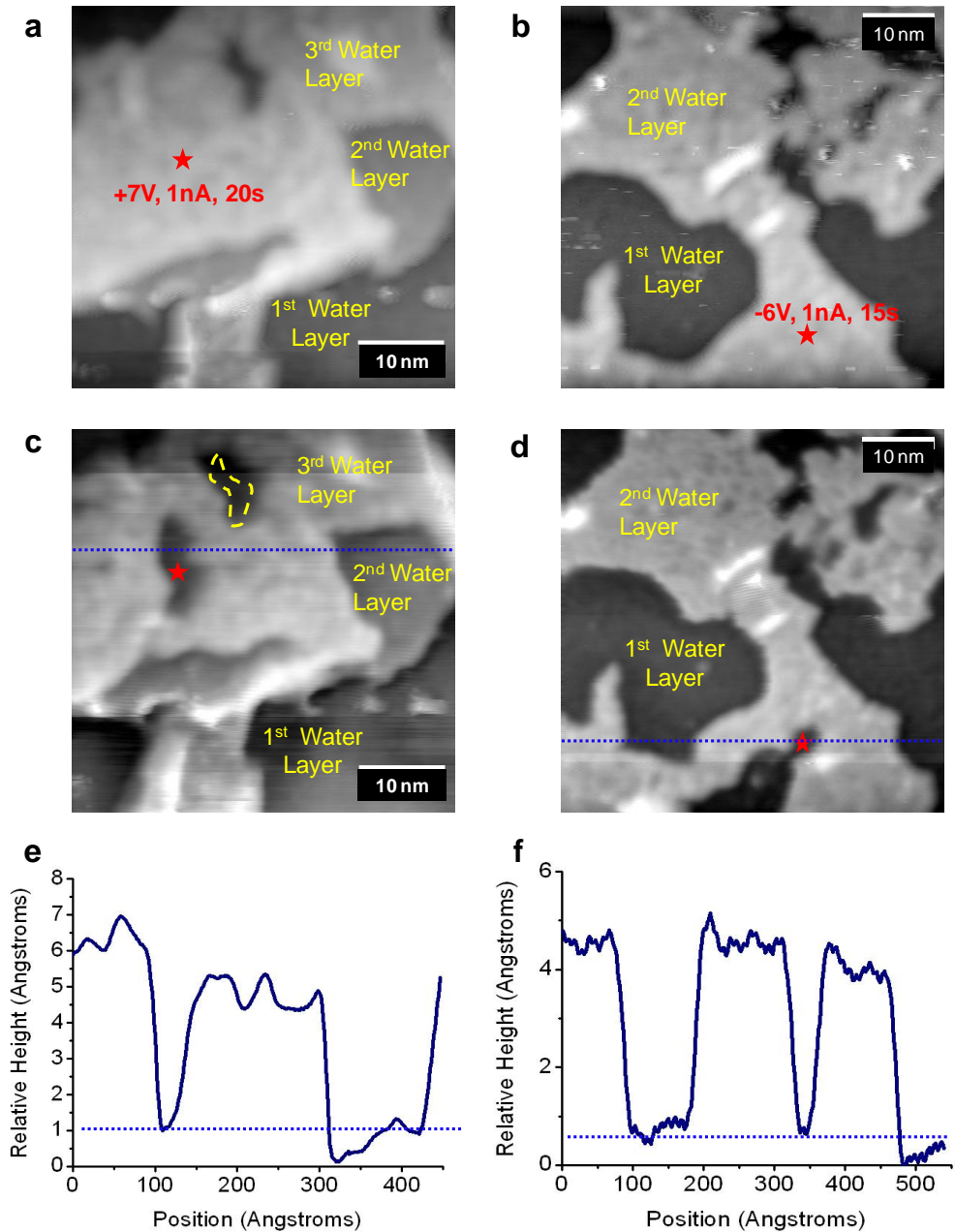

**Figure 5.** **(a)** STM topographic image of the third water layer before nano-manipulation at positive sample bias. **(b)** Topographic image of the second water layer before nano-manipulation at negative sample bias. **(c)** Topographic image of the same area in (a) after nano-manipulation at positive sample bias. The created pinhole is non-uniform, though it is localized to where the tip was centered. The dotted yellow line shows the outline of the original pinhole from (a), and we can see that this pinhole was also slightly enlarged after the manipulation. **(d)** Topographic image of the same area in (b) after nano-manipulation at negative sample bias. Similar to the positive bias case, the pinhole is again non-uniform, and appears to propagate in a random direction. **(e)** and **(f)** Height contours showing that for both the positive and negative bias case, the pinholes penetrate down to the water layer below. Scanning conditions were -0.35 V sample bias and 1 nA tunneling current.

**References**

(1) Verdaguer, A.; Sacha, G. M.; Bluhm, H.; Salmeron, M. *Chemical Reviews* **2006**, *106*, 1478-510.
(2) Thiel, P. A.; Madey, T. E. *Surface Science Reports* **1987**, *7*, 211-385.
(3) Brown, G. E. *Science* **2001**, *294*, 67-69.
(4) Guckenberger, R.; Heim, M.; Cevc, G.; Knapp, H. F.; Wiegräbe, W.; Hillebrand, A. *Science* **1994**, *266*, 1538-1540.
(5) Jang, C.; Adam, S.; Chen, J.-H.; Williams, E. D.; Das Sarma, S.; Fuhrer, M. S. *Physical Review Letters* **2008**, *101*, 146805.
(6) Shim, J.; Lui, C. H.; Ko, T. Y.; Yu, Y.-J.; Kim, P.; Heinz, T.; Ryu, S. *Nano Letters* **2012**, *12*, 648-654.
(7) Israelachvili, J. N.; Pashley, R. M. *Nature* **1982**, *300*, 341.
(8) Israelachvili, J. N.; Pashley, R. M. *Nature* **1983**, *306*, 249.
(9) Israelachvili, J. N.; Wennerström, H. *Nature* **1996**, *379*, 219.
(10) Raviv, U.; Klein, J. *Science* **2002**, *297*, 1540-1543.
(11) Granick, S.; Bae, S.; Kumar, S.; Yu, C. *Physics* **2010**, *3*, 1.
(12) Khan, S.; Matei, G.; Patil, S.; Hoffmann, P. *Physical Review Letters* **2010**, *105*, 106101-1 to 106101-4.
(13) Xu, K.; Cao, P.; Heath, J. R. *Science* **2010**, *329*, 1188-91.
(14) Severin, N.; Lange, P.; Sokolov, I. M.; Rabe, J. P. *Nano Letters* **2012**, *12*, 774.
(15) Geim, A. K.; Novoselov, K. S. *Nature Materials* **2007**, *6*, 183–191.
(16) Geim, A. K. *Science* **2009**, *324*, 1530-4.
(17) Zhang, Y.; Tan, Y.-W.; Stormer, H. L.; Kim, P. *Nature* **2005**, *438*, 201-4.
(18) Bolotin, K. I.; Sikes, K. J.; Jiang, Z.; Klima, M.; Fudenberg, G.; Hone, J.; Kim, P.; Stormer, H. L. *Solid State Communications* **2008**, *146*, 351-355.
(19) Novoselov, K. S.; Geim, A. K.; Morozov, S. V.; Jiang, D.; Zhang, Y.; Dubonos, S. V.; Grigorieva, I. V.; Firsov, A. A. *Science* **2004**, *306*, 666-9.
(20) Novoselov, K. S.; Geim, A. K.; Morozov, S. V.; Jiang, D.; Katsnelson, M. I.; Grigorieva, I. V.; Dubonos, S. V.; Firsov, A. A. *Nature* **2005**, *438*, 197-200.
(21) Du, X.; Skachko, I.; Barker, A.; Andrei, E. Y. *Nature Nanotechnology* **2008**, *3*, 491-5.
(22) Rutter, G. M.; Guisinger, N. P.; Crain, J. N.; Jarvis, E. A. A.; Stiles, M. D.; Li, T.; First, P. N.; Stroscio, J. A. *Physical Review B* **2007**, *76*, 235416.
(23) Schedin, F.; Geim, A. K.; Morozov, S. V.; Hill, E. W.; Blake, P.; Katsnelson, M. I.; Novoselov, K. S. *Nature Materials* **2007**, *6*, 652-5.
(24) Ishigami, M.; Chen, J. H.; Cullen, W. G.; Fuhrer, M. S.; Williams, E. D. *Nano Letters* **2007**, *7*, 1643-8.
(25) Ritter, K. A.; Lyding, J. W. *Nature Materials* **2009**, *8*, 235-242.
(26) He, K. T.; Koepke, J. C.; Barraza-Lopez, S.; Lyding, J. W. *Nano Letters* **2010**, *10*, 3446-52.
(27) Cao, P.; Xu, K.; Varghese, J. O.; Heath, J. R. *Journal of the American Chemical Society* **2011**, *133*, 2334-7.
(28) Mohanty, N.; Fahrenholtz, M.; Nagaraja, A.; Boyle, D.; Berry, V. *Nano Letters* **2011**, *11*, 1270-5.
(29) Cao, P.; Xu, K.; Varghese, J. O.; Heath, J. R. *Nano Letters* **2011**, *11*, 5581-6.

(30) Bunch, J. S.; Verbridge, S. S.; Alden, J. S.; van der Zande, A. M.; Parpia, J. M.; Craighead, H. G.; McEuen, P. L. *Nano Letters* **2008**, *8*, 2458-2462.
(31) Lui, C. H.; Liu, L.; Mak, K. F.; Flynn, G. W.; Heinz, T. F. *Nature* **2009**, *462*, 339-41.
(32) Lee, C.; Li, Q.; Kalb, W.; Liu, X.-Z.; Berger, H.; Carpick, R. W.; Hone, J. *Science* **2010**, *328*, 76-80.
(33) Wood, J. D.; Schmucker, S. W.; Lyons, A. S.; Pop, E.; Lyding, J. W. *Nano Letters* **2011**, *11*, 4547-4554.
(34) Li, X.; Cai, W.; An, J.; Kim, S.; Nah, J.; Yang, D.; Piner, R.; Velamakanni, A.; Jung, I.; Tutuc, E.; Banerjee, S. K.; Colombo, L.; Ruoff, R. S. *Science* **2009**, *324*, 1312-4.
(35) Zhang, W.; Wu, P.; Li, Z.; Yang, J. *The Journal of Physical Chemistry C* **2011**, *115*, 17782–17787.
(36) Bhaviripudi, S.; Jia, X.; Dresselhaus, M. S.; Kong, J. *Nano Letters* **2010**, *10*, 4128-33.
(37) Li, X.; Zhu, Y.; Cai, W.; Borysiak, M.; Han, B.; Chen, D.; Piner, R. D.; Colombo, L.; Ruoff, R. S. *Nano Letters* **2009**, *9*, 4359-63.
(38) Wu, Z.; Chen, Z.; Du, X.; Logan, J. M.; Sippel, J.; Nikolou, M.; Kamaras, K.; Reynolds, J. R.; Tanner, D. B.; Hebard, A. F.; Rinzler, A. G. *Science* **2004**, *305*, 1273-6.
(39) Nair, R. R.; Blake, P.; Grigorenko, A. N.; Novoselov, K. S.; Booth, T. J.; Stauber, T.; Peres, N. M. R.; Geim, A. K. *Science* **2008**, *320*, 1308.
(40) Ellison, M. D.; Good, A. P.; Kinnaman, C. S.; Padgett, N. E. *The Journal of Physical Chemistry. B* **2005**, *109*, 10640-6.
(41) Vedder, W.; McDonald, R. S. *The Journal of Chemical Physics* **1963**, *38*, 1583.
(42) Miranda, P.; Xu, L.; Shen, Y.; Salmeron, M. *Physical Review Letters* **1998**, *81*, 5876-5879.
(43) Kimmel, G. A.; Matthiesen, J.; Baer, M.; Mundy, C. J.; Petrik, N. G.; Smith, R. S.; Dohnálek, Z.; Kay, B. D. *Journal of the American Chemical Society* **2009**, *131*, 12838-44.
(44) Donadio, D.; Cicero, G.; Schwegler, E.; Sharma, M.; Galli, G. *The Journal of Physical Chemistry B* **2009**, *113*, 4170-5.
(45) Crupi, V.; Interdonato, S.; Longo, F.; Majolino, D.; Migliardo, P.; Venuti, V. *Journal of Raman Spectroscopy* **2008**, *39*, 244-249.
(46) Suk, J. W.; Kitt, A.; Magnuson, C. W.; Hao, Y.; Ahmed, S.; An, J.; Swan, A. K.; Goldberg, B. B.; Ruoff, R. S. *ACS Nano* **2011**, *5*, 6916-24.
(47) Ferrari, A. *Solid State Communications* **2007**, *143*, 47-57.
(48) Lenski, D. R.; Fuhrer, M. S. *Journal of Applied Physics* **2011**, *110*, 013720.
(49) Huang, M.; Yan, H.; Chen, C.; Song, D.; Heinz, T. F.; Hone, J. *Proc. Natl. Acad. Sci.* **2009**, *106*, 7304-8.
(50) Berciaud, S.; Ryu, S.; Brus, L. E.; Heinz, T. F. *Nano Letters* **2009**, *9*, 346-52.
(51) Liang, X.; Sperling, B. A.; Calizo, I.; Cheng, G.; Hacker, C. A.; Zhang, Q.; Obeng, Y.; Yan, K.; Peng, H.; Li, Q.; Zhu, X.; Yuan, H.; Hight Walker, A. R.; Liu, Z.; Peng, L.-M.; Richter, C. A. *ACS Nano* **2011**, *5*, 9144-9153.
(52) Malard, L. M.; Pimenta, M. A.; Dresselhaus, G.; Dresselhaus, M. S. *Physics Reports* **2009**, *473*, 51-87.
(53) Das, A.; Pisana, S.; Chakraborty, B.; Piscanec, S.; Saha, S. K.; Waghmare, U. V.; Novoselov, K. S.; Krishnamurthy, H. R.; Geim, A. K.; Ferrari, A. C.; Sood, A. K. *Nature Nanotechnology* **2008**, *3*, 210-5.
(54) Lin, Y.-C.; Lu, C.-C.; Yeh, C.-H.; Jin, C.; Suenaga, K.; Chiu, P.-W. *Nano Letters* **2012**, *12*, 414-419.

(55) Ni, Z.; Wang, Y.; Yu, T.; You, Y.; Shen, Z. *Physical Review B* **2008**, *77*, 113407.
(56) Cheng, L.; Fenter, P.; Nagy, K.; Schlegel, M.; Sturchio, N. *Physical Review Letters* **2001**, *87*, 156103.
(57) Pirkle, A.; Chan, J.; Venugopal, A.; Hinojos, D.; Magnuson, C. W.; McDonnell, S.; Colombo, L.; Vogel, E. M.; Ruoff, R. S.; Wallace, R. M. *Applied Physics Letters* **2011**, *99*, 122108.
(58) Chen, C.-C.; Bao, W.; Theiss, J.; Dames, C.; Lau, C. N.; Cronin, S. B. *Nano letters* **2009**, *9*, 4172-6.
(59) Huang, P. Y.; Ruiz-Vargas, C. S.; van der Zande, A. M.; Whitney, W. S.; Levendorf, M. P.; Kevek, J. W.; Garg, S.; Alden, J. S.; Hustedt, C. J.; Zhu, Y.; Park, J.; McEuen, P. L.; Muller, D. A. *Nature* **2011**, *469*, 389-92.
(60) Koepke, J. C.; Wood, J. D.; Estrada, D.; Ong, Z. Y.; Pop, E.; Lyding, J. W. *Submitted For Review* **2012**.
(61) Levy, N.; Burke, S. A.; Meaker, K. L.; Panlasigui, M.; Zettl, A.; Guinea, F.; Castro Neto, A. H.; Crommie, M. F. *Science* **2010**, *329*, 544-7.
(62) Sutter, P.; Sadowski, J. T.; Sutter, E. *Physical Review B* **2009**, *80*, 245411.
(63) Grantab, R.; Shenoy, V. B.; Ruoff, R. S. *Science* **2010**, *330*, 946-8.
(64) Park, S.-H.; Sposito, G. *Physical Review Letters* **2002**, *89*, 085501.
(65) Albrecht, P. M.; Lyding, J. W. *Applied Physics Letters* **2003**, *83*, 5029.
(66) HiPco Carbon Single Walled Carbon Nanotubes http://www.nanointegris.com/en/hipco (accessed Feb 15, 2012).
(67) Lyding, J. W.; Hubacek, J. S.; Tucker, J. R.; Abeln, G. C.; Shen, T.-C. *Applied Physics Letters* **1994**, *64*, 2010.
(68) Shen, T. C.; Wang, C.; Abeln, G. C.; Tucker, J. R.; Lyding, J. W.; Avouris, P.; Walkup, R. E. *Science* **1995**, *268*, 1590-2.
(69) Xu, Y.; He, K. T.; Schmucker, S. W.; Guo, Z.; Koepke, J. C.; Wood, J. D.; Lyding, J. W.; Aluru, N. R. *Nano Letters* **2011**, *11*, 2735-42.
(70) Xu, L.; Lio, A.; Hu, J.; Ogletree, D. F.; Salmeron, M. *The Journal of Physical Chemistry B* **1998**, *102*, 540-548.
(71) Morgenstern, K.; Rieder, K.-H. *Chemical Physics Letters* **2002**, *358*, 250-256.
(72) Gawronski, H.; Carrasco, J.; Michaelides, A.; Morgenstern, K. *Physical Review Letters* **2008**, *101*, 196101.
(73) Mehlhorn, M.; Gawronski, H.; Morgenstern, K. *Physical Review Letters* **2008**, *101*, 196101.
(74) Morgenstern, K.; Rieder, K.-H. *The Journal of Chemical Physics* **2002**, *116*, 5746.
(75) Maksymovych, P.; Dougherty, D.; Zhu, X.-Y.; Yates, J. *Physical Review Letters* **2007**, *99*.
(76) Odelius, M.; Bernasconi, M.; Parrinello, M. *Physical Review Letters* **1997**, *78*, 2855-2858.